\documentstyle[sprocl,epsfig,axodraw,graphics]{article}

\def\lsim{\mathrel{\rlap{\lower4pt\hbox{\hskip1pt$\sim$}}
    \raise1pt\hbox{$<$}}}         
\def\gsim{\mathrel{\rlap{\lower4pt\hbox{\hskip1pt$\sim$}}
    \raise1pt\hbox{$>$}}}         

\def\be{\begin{equation}}
\def\ee{\end{equation}}
\def\bq{\begin{eqnarray}}
\def\eq{\end{eqnarray}}

\def\Vec#1{\hbox{\boldmath$#1$\unboldmath}}
\def\slash{/\!\!\!}
\def\one{1 \!\! 1}
\def\D{{\mathrm d}}
\def\E{{\mathrm e}}
\def\I{{\mathrm i}}

\mathchardef\alpha="710B
\mathchardef\beta="710C
\mathchardef\gamma="710D
\mathchardef\delta="710E
\mathchardef\epsilon="710F
\mathchardef\zeta="7110
\mathchardef\eta="7111
\mathchardef\theta="7112
\mathchardef\iota="7113
\mathchardef\kappa="7114
\mathchardef\lambda="7115
\mathchardef\mu="7116
\mathchardef\nu="7117
\mathchardef\xi="7118
\mathchardef\pi="7119
\mathchardef\rho="711A
\mathchardef\sigma="711B
\mathchardef\tau="711C
\mathchardef\upsilon="711D
\mathchardef\phi="711E
\mathchardef\chi="711F
\mathchardef\psi="7120
\mathchardef\omega="7121
\mathchardef\varepsilon="7122
\mathchardef\vartheta="7123
\mathchardef\varpi="7124
\mathchardef\varrho="7125
\mathchardef\varsigma="7126
\mathchardef\varphi="7127
\mathchardef\nabla="7272
 
\font\dozeb=cmmib10 scaled \magstep1
\font\dozesyb=cmbsy10 scaled \magstep1
\font\dezb=cmmib10
\begin{document}

\title{TIME REVERSAL INVARIANCE \\ AND THE
TRANSVERSE SPIN STRUCTURE \\ OF HADRONS\footnote{Contribution 
to the Int. Work. on Nuclear Many-Body Problem and Sub-Nucleonic 
Degrees of Freedom in Nuclei (Changchun, China, July 2001) 
and to the Int. Conf. ``New Trends in High-Energy Physics'' 
(Yalta, Ukraine, September 2001).}
}

\author{Vincenzo Barone}

\address{Di.S.T.A., Universit{\`a} 
del Piemonte Orientale ``A.~Avogadro'', \\
INFN, Gruppo di Alessandria, 15100 Alessandria, Italy, \\
and Dipartimento di Fisica Teorica, \\
Universit{\`a} di Torino, 10125 Torino, Italy}

\maketitle\abstracts{The r{\^o}le of 
time reversal invariance in the phenomenology of 
transverse spin is discussed.}

\section{Introduction}
\label{intro}

 Time Reversal (TR) invariance is a fundamental constraint 
on many physical processes. It limits 
the admissible forms of structure functions, form factors, 
decay observables, etc. Acting on a momentum and 
spin eigenstate $\vert \Vec p, s \rangle$, the 
TR operator $T$ gives
\be
T \vert \Vec p, s_z  \rangle = (-1)^{s - s_z} 
\vert - \Vec p, - s_z \rangle \,, 
\label{TR1}
\ee
where $s$ is the particle's spin, $s_z$ its third 
component, 
and an irrelevant phase has been omitted.   
An important point, with far-reaching 
consequences,  
is that $T$ maps ``in'' states into ``out'' states:
$T : \vert {\rm in} \rangle 
\to \vert {\rm out} \rangle$. 

In what follows, we shall discuss the r{\^o}le 
that TR plays in the transverse spin structure 
of hadrons \cite{report}. Before entering into the 
subject, it is worth recalling that 
the implementation and the implications of TR invariance are 
sometimes rather subtle, as we are now going to show by a simple 
example \cite{Gasiorowicz}.  

\section{A pedagogical example}
\label{example}

Consider the decay of a particle of spin 
$\Vec s$ and zero momentum  into a state of spin $\Vec s'$ and 
momentum $\Vec p'$. 
Let $O( \Vec p'; \Vec s, \Vec s')$ be an 
observable. The expectation value of $O$ is 
\be
\langle O \rangle \sim \sum_{\Vec p', \Vec s, \Vec s'} 
O (\Vec p'; \Vec s, \Vec s') \, 
\vert \langle {\rm out}; \Vec p', \Vec s' 
\vert H \vert \Vec s \rangle \vert^2 \;, 
\label{ex1}
\ee
where $H$ is the interaction Hamiltonian responsible 
for the decay. 
Inserting a complete set of ``out'' states, 
labelled by the total angular momentum 
$J$ and its third component $m$, the matrix 
element in (\ref{ex1}) becomes
\be
\langle {\rm out}; \Vec p', \Vec s' 
\vert H \vert \Vec s \rangle =  
\sum_{Jm} 
\langle {\rm out}; \Vec p', \Vec s' 
\vert {\rm out}, J \, m  \rangle \, 
\langle {\rm out}, J \, m  \vert H \vert \Vec s \rangle \;. 
\label{ex1b}
\ee
It is easy to check, using the TR invariance of 
$H$, that the phase of 
$\langle {\rm out}, J \, m  \vert H \vert \Vec s \rangle$ is 
the phase shift for the channel with angular momentum $J$, 
that we call $\eta_J$. Thus eq.~(\ref{ex1b}) becomes 
\bq
\langle {\rm out}; \Vec p', \Vec s' 
\vert H \vert \Vec s \rangle
&=& 
\sum_{Jm} 
\langle {\rm out}; \Vec p', \Vec s' 
\vert {\rm out}, J \, m \rangle \, \E^{\I \eta_{J}}
\, \vert \langle {\rm out}, J \, m \vert H \vert \Vec s \rangle
\vert \nonumber \\
&\equiv& \sum_{J}  \E^{\I \eta_{J}}
\, M(J; \Vec p'; \Vec s, \Vec s') \;.  
\label{ex2}
\eq
In terms of $M$, eq.~(\ref{ex1}) reads
\be 
\langle O \rangle \sim \sum_{\Vec p', \Vec s, \Vec s'} 
O(\Vec p'; \Vec s, \Vec s') \, 
\sum_{J, J'} 
\E^{\I \, (\eta_{J} - \eta_{J'})} \, 
M(J; \Vec p'; \Vec s, \Vec s') 
\, M^*(J'; \Vec p'; \Vec s, \Vec s') \;. 
\label{ex2b}
\ee
TR invariance and the unitarity of the $S$-matrix, $S^{\dagger} S = \one$, 
imply (for the derivation of this result 
see the book by Gasiorowicz \cite{Gasiorowicz}) 
\be
 M^*(J; \Vec p'; \Vec s, \Vec s')
=  M (J; - \Vec p'; - \Vec s, -\Vec s') \;. 
\label{ex7}
\ee
Suppose now that $O$ is {\em odd under TR}, that is 
\be
O (\Vec p'; \Vec s, \Vec s') = 
- O (- \Vec p'; -\Vec s, -\Vec s') \;. 
\label{ex8}
\ee
Then, 
with the help of (\ref{ex7}),  
eq.~(\ref{ex2b}) gives 
\bq
\langle O \rangle 
& \sim & 
\sum_{\Vec p', \Vec s, \Vec s'}
\sum_{J, J'} \sin (\eta_{J} - \eta_{J'}) 
\, O(\Vec p'; \Vec s, \Vec s') \nonumber \\
& & \times \,  
M(J; \Vec p'; \Vec s, \Vec s') 
\, M(J'; - \Vec p'; - \Vec s, - \Vec s'). 
\label{ex9}
\eq
This shows that, 
in spite of the fact that $O$ is $T$-odd, its 
expectation value does not vanish if $\sin (\eta_{J} 
- \eta_{J'}) \neq 0$, which may happen in 
presence of final state interactions that generate 
non-trivial phase differences between the various reaction
channels. 

Thus, the important lesson is that when final-state 
(or initial-state) non-trivial effects are at work, 
observables which are {\em na{\"\i}vely} $T$-odd according 
to their structure in terms of spins and momenta, 
may give rise to non-zero measurable quantities, 
without really violating TR invariance. 

A noteworthy example is provided by pion-nucleon scattering.  
Although the correlation 
\be
(\Vec P_{\pi} \times \Vec P_{N}) \cdot 
\Vec S_{N} \;, 
\label{ex10}
\ee 
is $T$-odd in the sense of (\ref{ex8}), its 
vacuum expectation value is known to be non zero.

\section{Semi-inclusive leptoproduction}
\label{sidis}

\begin{figure}[t]
\begin{center}
\begin{picture}(300,190)(0,0)
\SetWidth{3}
\ArrowLine(70,25)(120,40)
\ArrowLine(180,40)(230,25)
\SetWidth{0.5}
\ArrowLine(120,60)(120,90)
\ArrowLine(180,90)(180,60)
\ArrowLine(120,90)(120,120)
\ArrowLine(180,120)(180,90)
\Photon(80,90)(120,90){3}{4.5}
\Photon(220,90)(180,90){3}{4.5}
\GOval(150,50)(20,40)(0){0.5}
\SetWidth{2}
\ArrowLine(120,142)(130,170)
\ArrowLine(170,170)(180,142)
\SetWidth{0.5}
\GOval(150,130)(20,40)(0){0.5}
\LongArrow(75,100)(85,100)
\LongArrow(215,100)(225,100)
\Text(65,25)[r]{$P, S$}
\Text(235,25)[l]{$P, S$}
\Text(80,80)[]{$q$}
\Text(220,80)[]{$q$}
\Text(125,170)[r]{$P_h$}
\Text(175,170)[l]{$P_h$}
\Text(115,75)[r]{$k, s$}
\Text(185,75)[l]{$k, s$}
\Text(115,105)[r]{$\kappa, s'$}
\Text(185,105)[l]{$\kappa, s'$}
\Text(150,130)[]{\LARGE $\Xi$}
\Text(150,50)[]{\LARGE $\Phi$}
\DashLine(150,20)(150,35){4}
\DashLine(150,65)(150,115){4}
\DashLine(150,145)(150,165){4}
\end{picture}
\end{center}

\caption{Diagram contributing to semi-inclusive DIS 
in leading order QCD and leading twist.}
\label{handbag2}
\end{figure}
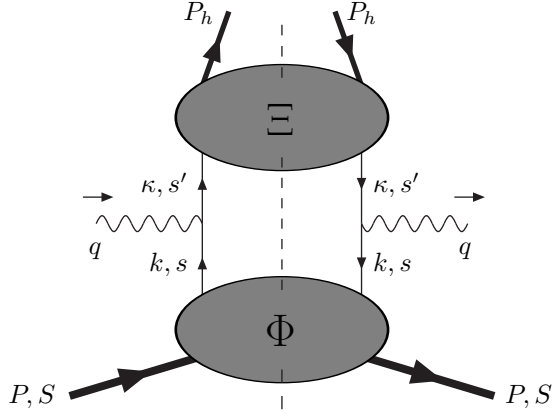

Our  prototype  process will be 
 semi-inclusive deep inelastic scattering 
off a transversely polarised target, 
\be
l (\ell) \, + \, N^{\uparrow} (P) \rightarrow l' (\ell') \, + \, h (P_h) 
\, + \, X (P_X) \;,  
\label{sidis0}
\ee
whose cross section reads
\be
\frac{\D \sigma}{ \D x \,  \D 
y \,  \D z \, \D^2 \Vec P_{h \perp}} = \frac{\pi 
\alpha_{\rm em}^2 \, y}{2 \, Q^4 \, z}  
\, L_{\mu \nu} W^{\mu \nu} \;.  
\label{sidis9}
\ee
Here $L_{\mu \nu}$ is the usual leptonic tensor 
of DIS, whereas $W^{\mu \nu}$ is the hadronic 
tensor, which is given by, in leading order QCD 
and leading twist (see Fig.~\ref{handbag2}
for notations)
\bq
& & 
W^{\mu \nu} = 
\sum_a e_a^2  \, 
\int \! \frac{\D k^+ \, \D k^- \, \D^2 \Vec k_T}{(2 \pi)^4}  
\int \! \frac{\D \kappa^+ \, \D \kappa^- \, 
\D^2 \Vec \kappa_T }{(2 \pi)^4} 
\nonumber \\
 & & \hspace{0.5cm} \times \, 
\delta (k^+ - x P^+) \, \delta (k^-  - P_h^- /z)
\, \delta^2 (\Vec k_T + \Vec q_T - {\Vec \kappa}_T) \, 
  {\rm Tr} \, [\Phi  \, \gamma^{\mu} \, \Xi  
\gamma^{\nu}],  
\label{sidis21}
\eq
with $z = P \cdot P_h/ P \cdot q$.

The quark structure of hadrons is incorporated into 
the correlation matrix $\Phi$ and the decay matrix 
$\Xi$. These matrices are defined as ($i,j$ are Dirac indices)
\bq
\Phi_{ij}(k, P, S) &=& 
\int \D^4 \xi \, 
\E^{i k \cdot \xi} \, 
\langle P S \vert \overline \psi_j(0) \psi_i(\xi) \vert P S \rangle \;. 
\label{df1} \\
\Xi_{i j}(\kappa; P_h, S_h) &=&
\sum_{X} 
\int \! \D^4 \xi \, \E^{i \kappa \cdot \xi} 
 \langle 0 \vert  \psi_i(\xi) \vert P_h S_h, X \rangle 
\langle P_h S_h, X \vert 
\overline \psi_j (0) \vert 0 \rangle.  
\label{frag1}
\eq
$\Phi$ contains the
{\em distribution functions}; $\Xi$ contains the
{\em fragmentation functions}.

The $T$-odd correlations 
we shall be interested in  are similar to (\ref{ex10}), 
but involve the transverse momenta of quarks. They are 
\be
 (\Vec k_{\perp} \times \Vec P) \cdot \Vec S \;, 
\;\;
(\Vec k_{\perp} \times \Vec P) \cdot \Vec s \;, 
\label{corr2} 
\ee
\be
 (\Vec \kappa_{\perp} \times \Vec P_h) \cdot \Vec s'\;.  
\label{corr3}
\ee
Note that, while the first two correlations 
involve the momentum and/or the spin 
of the quark 
inside the target hadron, the third correlation 
involves the momentum 
and the spin of the fragmenting quark. According to 
the general discussion of Sect.~(\ref{example}), 
the correlations (\ref{corr2})  may give rise to 
observable effects due to some {\em initial-state 
interactions}, 
whereas (\ref{corr3}) may be observable 
due to {\em final-state 
interactions}.

\section{TR invariance and quark distribution 
functions}

Time reversal invariance translates into the 
following condition \cite{Mulders} on $\Phi$, 
\be
\Phi^{*}(k, P, S) = \gamma^5 {\mathcal C} \,  
\Phi(\widetilde {k}, \widetilde {P},
\widetilde {S}) \, {\mathcal C}^{\dagger} \gamma^5 ,  
\label{df1d} 
\ee
where ${\mathcal C} = \I \gamma^2 \gamma^0$ and the tilde 
four-vectors are defined as $\widetilde {k}^{\mu} = (k^0, 
- \Vec k)$. This relation is obtained by using 
\be
T \, \psi_a(\xi) \, T^{\dagger} = - \I \gamma_5 {\mathcal C} 
\psi_a (- \widetilde{\xi}) 
\label{tpsi}
\ee 
and $T \vert P S \rangle = (-1)^{S - S_z} \vert \widetilde{P} 
\widetilde{S} \rangle$.  

If we ignore (or integrate over) the transverse motion of quarks,  
the TR constraint (\ref{df1d}) has no effect on the structure of 
$\Phi$ at leading twist. In this case, the integrated 
quark--quark correlation 
matrix $\Phi$ 
\be
\Phi_{i j}(x) = \int \frac{\D^4 k}{(2 \pi)^4} 
\, \Phi_{ij}(k, P, S) \, \delta  \left (x - \frac{k^+}{P^+} \right ) 
\label{df22}
\ee
reads 
\be
\Phi (x) = 
\frac{1}{2} \, 
\left \{ f(x)  \, \slash P + \lambda_N \, \Delta f(x) \, 
\gamma_5  \, \slash P  + \Delta_T f(x) \, \slash P \, \gamma_5 \, 
\slash S_{\perp} \right \} \;, 
\label{df23}
\ee
where $f(x)$, $\Delta f(x)$ and $\Delta_T f(x)$ are 
the unpolarised, the helicity and the transversity 
distributions, respectively. 

At twist 3, on the contrary, the TR property (\ref{df1d}) 
does play a r{\^o}le. It forbids, for instance, a pseudoscalar 
term (which does not contribute to leading twist as it is 
suppressed by a factor $1/P^+$ in the infinite momentum frame). 
If we relax the condition (\ref{df1d}) 
-- for a justification, see below --, we get a $T$-odd  
twist--3 correlation matrix, which contains three 
distribution functions \cite{Boer1}
\be
\left. \Phi (x) \right \vert_{TR-{\rm odd}} =  
\frac{M}{2} \, 
\left \{ f_T(x) \, \varepsilon_{\perp}^{\mu \nu} S_{\perp \mu} \gamma_{\nu}
 - \I \lambda_N \, e_L(x) \, \gamma_5 \,  
+ \frac{\I}{2} \, h(x) \,  
[ \slash p, \slash n ] \right \}, 
\label{higher7}
\ee
where $\varepsilon_{\perp}^{\mu \nu} = 
\varepsilon^{\mu \nu \rho \sigma} P_{\rho} q_{\sigma}/ 
P \cdot q$. 

As shown by Boer, Mulders and Teryaev \cite{Boer3}, 
there is no need to invoke initial-state interactions
to justify 
the existence of  
$f_T(x)$, $e_L(x)$ and $h(x)$. These arise as 
{\em effective} distributions 
 related to 
the multiparton densities which contribute 
to higher twists. The point is that the
 twist-3 hadronic tensor contains, besides
$\Phi$, a quark-quark-gluon correlation 
matrix which may have no definite 
behavior under TR. The condition (\ref{df1d}) 
does not apply to it and $T$-odd distributions  
are allowed. We emphasize that this mechanism only works 
at higher twists.

\subsection{$T$-odd couple: $\Delta_0^T f$ and 
$\Delta_T^0 f$}

Let us return to leading twist.
When the transverse motion of quarks inside the target 
is taken into account, the structure of $\Phi$ is more 
complicated than (\ref{df23}), and the condition (\ref{df1d}) 
becomes truly restrictive. In particular, it forbids 
terms in $\Phi$ of the form 
\bq
& &  \varepsilon^{\mu \nu \rho \sigma} \, \gamma_{\mu} P_{\nu} 
k_{\perp \rho} S_{\perp \sigma} , 
\label{todd1} \\
& & 
 \varepsilon^{\mu \nu \rho \sigma} \, \gamma_5 \sigma_{\mu \nu} P_{\rho}
k_{\perp \sigma}, 
\label{todd2}
\eq
which give rise to two $\Vec k_{\perp}$-dependent 
TR--odd distribution functions, that we 
call $\Delta_0^T f$ and 
$\Delta_T^0 f$. 
 The former 
 is related to the number density of unpolarised quarks
in a transversely polarised nucleon; the latter  
measures the transverse
polarisation of quarks in an unpolarised hadron. 
If we call ${\mathcal P}_{a/p} (x, \Vec k_{\perp})$ 
the probability to find a quark $a$ with momentum fraction $x$ 
and transverse momentum $\Vec k_{\perp}$ in the target proton, 
we have \cite{report} ($\vert P, \pm \rangle$ 
are the momentum-helicity eigenstates of 
the proton) 
\bq
& &  {\mathcal P}_{a/p^{\uparrow}} (x,   \Vec{k}_\perp) -
  {\mathcal P}_{a/p^{\uparrow}} (x, - \Vec{k}_\perp)
  \nonumber \\   
& & = {\rm Im} \; \int{\D y^- \, \D^2 \Vec y_\perp\over 2(2 \pi)^3}
{\rm e}^{-\I x P^+ y^- + \I  \Vec k_\perp\cdot \Vec y_\perp} \,
 \langle P, - \vert \overline{\psi}_a(0,y^-, y_\perp)
{\gamma^+}\psi_a(0)\vert P, + \rangle  \nonumber \\
& & \equiv 
\frac{(\Vec k_{\perp} \times \Vec P) \cdot \Vec S_{\perp}}{\vert 
\Vec k_{\perp} \times \Vec P \vert \, \vert \Vec 
S_{\perp} \vert} \, 
\Delta_0^T f(x, \Vec k_{\perp}^2) , 
\label{todd3}
\eq 
and 
\bq
& &  {\mathcal P}_{a^{\uparrow}/p}   (x, \Vec{k}_\perp) -
  {\mathcal P}_{a^{\downarrow}/p} (x, \Vec{k}_\perp) 
\nonumber \\
& & = 
 \int{\D y^- \, \D^2 \Vec y_\perp\over 2(2 \pi)^3}
{\rm e}^{-\I x P^+ y^- + \I \Vec k_\perp\cdot \Vec y_\perp} 
\,  \langle P \vert \overline{\psi}_a(0,y^-,y_\perp)
{\I \sigma^{2 +} \gamma_5 }\psi_a(0)\vert P \rangle \nonumber \\
& & \equiv 
\frac{(\Vec k_{\perp} \times \Vec P) \cdot \Vec s_{\perp}}{\vert 
\Vec k_{\perp} \times \Vec P \vert \, \vert \Vec 
s_{\perp} \vert} \, 
\Delta_T^0 f(x, \Vec k_{\perp}^2). 
\label{todd4}
\eq 
(The transverse polarisation of the quarks and of the proton 
is denoted by arrows and  
assumed to be directed along the $y$-axis.)

In the literature \cite{Boer1} one often finds two other 
functions, $f_{1T}^\perp$ and $h_1^\perp$, related to  
$\Delta_0^T f$ and $\Delta_T^0 f$ by  
\bq
\Delta_0^T f(x, \Vec k_{\perp}^2) &=& 
 -2 \, \frac{| \Vec{k}_\perp |}{M} \, 
  f_{1T}^\perp(x, \Vec{k}_\perp^2) , \label{f1t} \\
\Delta_T^0 f(x, \Vec k_{\perp}^2) &=& 
- \frac{| \Vec{k}_\perp |}{M} \,
  \, h_1^\perp(x, \Vec{k}_\perp^2) .
\label{h1perp}
\eq


The distribution $\Delta_0^T f$ 
was first introduced by Sivers \cite{Sivers}  
and its phenomenological applications were investigated  
by several authors \cite{Mauro,Boer1,Boglione}. 
 The distribution 
$\Delta_T^0 f$ was studied by Boer 
and Mulders \cite{Boer1,Boer2}.
Their $T$-odd character can be checked by direct 
inspection. 
Using the standard action of TR on quark fields, {\it i.e.} 
eq.~(\ref{tpsi}), 
it is easy to show that  
the matrix elements in (\ref{todd3},\ref{todd4}) 
 change sign 
under $T$, and therefore the corresponding 
distributions must vanish 
(this was first pointed out 
by Collins \cite{Collins}).

Let us now comment on the physical 
meaning of the distributions we have just introduced. 
One may legitimately wonder whether $T$-odd quantities, such as 
$\Delta_0^T f$ and $\Delta_T^0 f$, 
make any sense at all. 
In order to justify the existence of 
these quantities, their proponents \cite{Mauro}
advocate initial-state hadronic interactions, 
which prevent implementation  of time-reversal invariance 
via the condition (\ref{df1d}). 
The idea is that 
the colliding particles interact strongly with 
non-trivial relative phases. 
If this is correct, $\Delta_0^T f$ and $\Delta_T^0 f$
should only be observable  in 
reactions involving two initial hadrons (Drell-Yan processes, 
hadron production in proton-proton collisions, etc.), 
not in leptoproduction.

The Sivers function $\Delta_0^T f$ may account for the 
observed single-spin asymmetry in transversely polarised 
pion hadroproduction (the so-called 
Sivers effect, see Sect.~\ref{phenom}). The 
distribution $\Delta_T^0 f$ has been used by Boer \cite{Boer2}
to explain, at leading--twist level, an anomalously  
large $\cos 2 \phi$ term in the unpolarised 
Drell--Yan cross section.
Introducing initial--state $T$-odd effects, 
the unpolarised Drell--Yan cross section acquires indeed a 
$\cos 2 \phi$ contribution proportional to the product  
$\Delta_T^0 f \times \Delta_T^0 \overline{f}$. 

The real difficulty about the $T$-odd distributions is that  
no initial-state interaction mechanism 
is known which can produce such things as 
$\Delta_0^T f$ and $\Delta_T^0 f$. Therefore 
 their existence is 
-- to say the least -- highly questionable.

A different way of looking at the $T$-odd distributions 
is presently under investigation \cite{noi}.
It is based on a ``non-standard'' time reversal for 
particle multiplets \cite{Weinberg}, 
which turns out to be a good symmetry in chiral quark models
of the nucleon. 
Equation~(\ref{tpsi}) is replaced by  
\be
T \, \psi_a(\xi) \, T^{\dagger} = - \I (\tau_2)_{ab} 
\gamma_5 {\mathcal C} 
\psi_b (- \widetilde{\xi}),   
\label{tpsib}
\ee 
where $\Vec \tau$ is the isospin operator. The time 
inversion in (\ref{tpsib}) relates different 
components of the flavor multiplet and consequently 
 (\ref{todd3}) and (\ref{todd4}) 
do not vanish any longer once TR invariance is imposed  
(via (\ref{tpsib})).
 If this mechanism 
is correct, the $T$-odd distributions  should also be observable  
 in semi-inclusive leptoproduction. 
A conclusive statement on the matter will only be made by  
experiments.


\section{TR invariance and quark fragmentation 
functions}
\label{fragment}

In the fragmentation process
one cannot 
na{\"\i}vely impose a condition similar to (\ref{df1d}), that is 
\be
\Xi^*(\kappa, P_h, S_h) = \gamma^5 {\mathcal C}  
\, \Xi(\widetilde {\kappa}, \widetilde {P}_h, \widetilde {S}_h) \, 
{\mathcal C}^{\dagger} \gamma^5. 
\label{frag7bis}
\ee 
In the derivation of (\ref{df1d}) the 
simple transformation property of the nucleon 
 state $\vert P S \rangle$ under $T$ is crucial. However,  
   $\Xi$ contains $\vert P_h S_h, X \rangle$ 
which are ``out'' states. 
  Under time reversal they 
do not simply invert their momenta and spins 
but transform 
into ``in'' states
\be
T \,\vert P_h S_h, X; {\rm out} \rangle \propto
\vert \widetilde {P}_h \widetilde {S}_h, \widetilde {X}; 
{\rm in} \rangle.  
\label{frag7b}
\ee
These may differ non trivially from $\vert \widetilde {P}_h 
\widetilde {S}_h, \widetilde{X}; {\rm out} \rangle$, due to  
final-state interactions, which can generate 
relative phases between the various 
channels of the $\vert {\rm in} \rangle 
\to \vert {\rm out} \rangle$ transition. 
As a consequence, a term in $\Xi$ of the form 
\be
 \varepsilon^{\mu \nu \rho \sigma} \, \gamma_5 \sigma_{\mu \nu} P_{\rho}
\kappa_{T \sigma} 
\label{toddfrag1}
\ee
is not forbidden by time reversal invariance, 
 and generates 
a $T$-odd fragmentation function, $\Delta_T^0 D$, 
given by 
\be
{\mathcal N}_{h/a \uparrow} (z, \Vec \kappa'_{T}) 
- {\mathcal N}_{h/a \downarrow} (z, \Vec \kappa'_{T}) 
= \frac{( \Vec \kappa_T \times \Vec P_h) 
\cdot \Vec s'}{\vert \Vec \kappa_T \times \Vec P_h \vert 
\, \vert \Vec s' \vert} \, \Delta D_T^0 
(z, {{\Vec \kappa}'_T}^2), 
\label{kappaT11} 
\ee
where $\Vec \kappa'_T = - z \Vec \kappa_T$. 
$\Delta_T^0 D$
 is responsible for the  
the so--called 
Collins effect\cite{Collins,Mauro2}, 
{\it i.e.}, non-zero
azimuthal asymmetries  in single--inclusive
production of unpolarised hadrons at leading twist. 
In Mulders' classification \cite{Mulders}, 
a function $H_{1 \perp}$ 
is introduced, which is related to $\Delta_T^0 D$
by 
\be
\Delta_T^0 D (z, {{\Vec \kappa}'_T}^2) 
= \frac{\vert \Vec \kappa_T \vert}{M_h} \, 
H_{1 \perp} (z, {{\Vec \kappa}'_T}^2) \;. 
\label{kappaT11b}
\ee
 The factor in front of $\Delta_T^0 D$
in (\ref{kappaT11}) is the sine of the azimuthal angle 
between the spin vector and the momentum of the fragmenting quark, 
the so-called ``Collins angle'' $\Phi_C$.

Just to see in practice 
 how the $T$-odd fragmentation function 
$\Delta_T^0 D$ may
arise from non trivial final--state interactions,  let 
us assume that a quark
 fragments into an unpolarised hadron,  
leaving, as  a remnant, a pointlike scalar diquark \cite{Bianconi}. 
If we describe the hadron $h$ by a plane wave,  
\be
\psi_h (x) \sim u(P_h) \, \E^{i \, P_h \cdot x}, 
\label{kappaT15}
\ee
it is easy to show that the fragmentation matrix $\Xi$ is 
\bq
\Xi(\kappa, P_h) &\sim& 
\frac{- \I}{\slash \kappa - m} \, u(P_h) \, \overline{u}(P_h)
\, \frac{\I}{\slash \kappa - m} \nonumber \\ 
 &\sim&  
\frac{\slash \kappa + m}{\kappa^2 - m^2} 
\, (\slash P_h + M_h ) \, 
\frac{\slash \kappa + m}{\kappa^2 - m^2}, 
\label{kappaT16}
\eq
where $m$ is the quark mass and we have omitted some inessential factors.
From (\ref{kappaT16}) we cannot extract a term  of the type
(\ref{toddfrag1}) (hence, producing $\Delta_T^0 D$). 
Let us now suppose that a residual interaction of $h$ with 
the intermediate state generates a phase in the hadron 
wave function. If, for instance, we make in (\ref{kappaT16}) 
the replacement (assuming only two fragmentation channels)
\be
u(P_h) \rightarrow u(P_h) + \E^{i \chi} \, \slash \kappa
\, u(P_h), 
\label{kappaT17}
\ee
by a little algebra one can show that $\Xi$ acquires 
a term 
\be
\frac{M_h}{\kappa^2 - m^2} \, \sin \chi \; 
 \varepsilon^{\mu \nu \rho \sigma} \, \gamma_5 \sigma_{\mu \nu} P_{\rho}
k_{\perp \sigma} . 
\label{toddfrag1b}
\ee
Therefore, if the interference between the fragmentation 
channels produces a non--zero phase $\chi$,  a $T$-odd 
contribution may appear. 

Bacchetta {\it et al.} \cite{Bacchetta} have recently shown 
that in a chiral quark model with pseudoscalar 
couplings the Collins fragmentation function is generated
by one-loop 
self-energy and vertex correction diagrams.  

However, as argued by Jaffe, Jin and Tang \cite{Jaffe}, 
the proliferation of channels 
and the sum over $X$ 
might ultimately lead to an overall cancellation 
of the relative phases between the channels, 
and to the vanishing 
of the Collins function. If this is true (and 
only experiments will provide a definite answer), then,   
in order to observe a $T$-odd 
correlation in 
the fragmentation process, one should rather consider 
correlated variables belonging to physical 
particles with known interactions. 
This suggests to return to a 
quantity {\em exactly} like  (\ref{ex10}), namely 
\be
(\Vec P_1 \times \Vec P_2) \cdot \Vec S \;, 
\label{twohad1}
\ee
where $\Vec P_1$ and $\Vec P_2$ are the momenta of two 
final-state hadrons. 
Two-hadron leptoproduction,
\be
l (\ell) \, + \, N^{\uparrow} (P) \rightarrow l' (\ell') \, + \, h_1 (P_1) 
\, + \, h_2 (P_2) \, + \, 
X (P_X) .
\label{twopart1}
\ee
has been proposed \cite{Jaffe} and studied \cite{Jaffe,Bianconi}
as a potential source of information  
about transversity and $T$-odd correlations. In the decay 
matrix the term proportional to (\ref{twohad1}) contains 
a fragmentation function $\Delta_T I(z, \xi, M_h^2)$, 
where $M_h = P_h^2 \equiv (P_1 + P_2)^2$ and $\xi = P \cdot P_1/ 
P \cdot P_h$. In the specific case of 
$\pi^+ \pi^-$ production, $\Delta_T I$ arises from 
the interference between the $s$- and $p$-wave of the 
pion system, which is known from the experiment to be 
non zero.

\section{Single-spin asymmetries}
\label{phenom}

As they involve transversely polarised quarks 
in unpolarised hadrons, or {\it viceversa}, 
the $T$-odd distribution and fragmentation functions 
may give rise to single-spin transverse asymmetries in 
lepto- and hadroproduction. 

\begin{figure}[t]
\begin{center}
\mbox{\epsfig{file=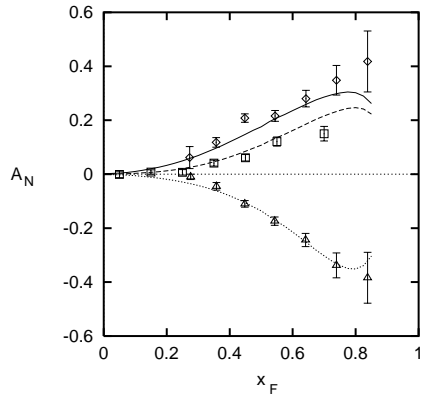,width=5cm,angle=-90}}
\end{center}
\caption{Fit of the data on $A_T^{\pi}$ for the 
process $p^{\uparrow} p \to \pi X$ assuming that only 
Collins effect is active (Ref.~13); the upper, middle, 
and lower sets of data and curves refer 
to $\pi^+$, $\pi^0$ and $\pi^-$, respectively.}
\label{pion}
\end{figure}

In 1991, the E704 experiment at Fermilab \cite{E704}
discovered a sizeable single-spin asymmetry in inclusive 
pion hadroproduction with a transversely polarised 
proton (Fig.~\ref{pion}). This result, which cannot be 
explained  by perturbative QCD in leading 
twist, may be attributed either to the $T$-odd 
distribution function $\Delta_0^T f$ (Sivers effect) or to 
the $T$-odd fragmentation function $\Delta_T^0 D$ 
(Collins effect)\footnote{An explanation in terms 
of $\Delta_T^0 f$ is also possible.}.
In the former case one has 
\be 
\D \sigma^{\uparrow} - 
\D \sigma^{\downarrow} \sim 
\sum_{abc} 
\int \D x_a \, \D x_b  \, \D^2 \Vec k_T \;
 \Delta_0^T f_a(x_a, \Vec k_T^2) \, f_b(x_b) \, 
\D \hat \sigma \, 
  D_{c}(z),
\label{hh17}
\ee
where $\D \hat \sigma$ is a partonic cross section and 
$D(z)$ is the familiar unpolarised fragmentation function.
The Collins mechanism gives 
\be
\D \sigma^{\uparrow} - 
\D \sigma^{\downarrow} \sim
 \sum_{abc} 
\int \D x_a \, \D x_b  \, \D^2 \Vec \kappa_T \;
 \Delta_T f_a(x_a) \,  f_b(x_b) \, 
\Delta_{TT} \hat \sigma \, 
 \Delta_T^0 D_{c}(z, \Vec \kappa_T^2),
\label{hh18}
\ee
where $\Delta_{TT} \hat \sigma$ is a partonic double-spin 
asymmetry.

Fig.~\ref{pion} shows the asymmetry predicted by Anselmino {\it et al.} 
 \cite{Mauro2} using eq.~(\ref{hh18}) and 
a simple  parametrisation of the Collins function. 
An equally good description is obtained by means of eq.~(\ref{hh17}).
 An alternative theoretical 
picture to Collins and Sivers mechanisms is 
 based on higher-twist, non $T$-odd,
 distribution and fragmentation functions \cite{Qiu}. 
The investigation of the $p_T$ dependence of the process 
would clearly help to distinguish between leading-twist 
and higher-twist effects. 

\begin{figure}[t]
\begin{center}
\mbox{\epsfig{figure=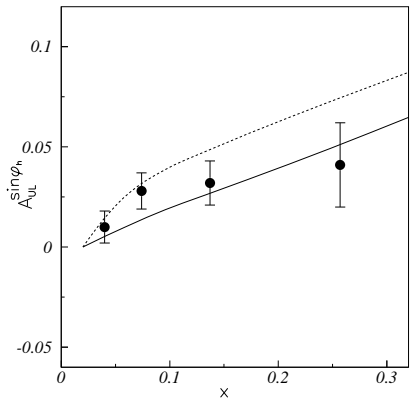,width=6cm}}
\end{center}
\caption{The HERMES data on the transverse single-spin asymmetry 
in pion leptoproduction with the fits of De Sanctis {\it et al.}
(Ref.~21).
The solid line corresponds to $\Delta_T f = \Delta f$, the dashed 
line corresponds to saturation of the Soffer inequality, 
$\vert \Delta_T f \vert = (f + \Delta f)/2$.}
\label{fig_hermes}
\end{figure}

The first (preliminary) measurements of single-spin transverse asymmetries 
in pion leptoproduction have been presented two years ago 
by HERMES \cite{HERMES} and SMC \cite{SMC}. The HERMES result 
is shown in Fig.~\ref{fig_hermes}. The $\sin \phi$ contribution to 
the asymmetry (where $\phi$ is the 
azimuthal angle of the produced pion) has the form
\be
A^T \sim \Delta_T f (x) \, \Delta_T^0 D(z, \Vec P_{\pi \perp}^2) 
\, \sin \phi \;.  
\label{atlepto} 
\ee
A fit \cite{desanctis} of the data, based on (\ref{atlepto}) and 
on two different assumptions for the transversity 
distribution $\Delta_T f$, is  
displayed in Fig.~\ref{fig_hermes}. As one can see, the agreement is 
fairly good (the HERMES data are also well 
reproduced by a light-cone quark model \cite{Ma}). From the result on $A^T$
one can derive a lower bound for the quark 
analysing power $\Delta_T^0 D/D$, namely \cite{Mauro3}
\be
\frac{\vert \Delta_T^0 D \vert}{D} 
\gsim 0.20 , 
\;\; \;
 z \ge 0.2 \;. 
\ee

Concluding this phenomenological account, 
it is fair to say that the present scarcity 
of data, their uncertainties and our ignorance of most 
of the quantities involved in the process make the entire 
matter still rather vague. More, and more precise, measurements 
are clearly needed to get a better understanding.

\section{Perspectives}

In the next few years, the upgraded HERMES experiment at 
HERA and the COMPASS experiment at CERN (which 
upgrades SMC) will provide more accurate measurements 
of the single-spin transverse asymmetry in semi-inclusive pion 
production and, hopefully, the first measurements 
of the transverse asymmetry in two-hadron production. 
This should allow us to achieve two goals: {\it i)}
to extract for the first time the transversity distributions 
of the nucleon and {\it ii)} to reveal in a clear way 
possible $T$-odd effects. In order to disentagle 
mechanisms of different nature giving rise to single-spin asymmetries,  
the study of the $Q^2$ dependence of the processes
will be crucial.

\section*{Acknowledgments}

I would like to thank the organizers of the 
Changchun workshop and of the Yalta conference  
for their kind invitation.
It is a pleasure to acknowledge the
hospitality of Jilin University, 
Peking University, the Institute 
of High Energy Physics of the Chinese 
Academy of Sciences and the Bogolyubov Institute 
for Theoretical Physics of the Ukrainian Academy of Sciences.
 I am grateful to 
Mauro Anselmino, Alessandro Drago and Francesco Murgia
 for collaboration on the subject of this paper, and to  
Bo-Qiang Ma, Guo-Mo Zeng, Bing-Song Zou, Oleg Teryaev and 
Laszlo Jenkovszky for useful 
discussions.



\section*{References}
\begin{thebibliography}{99}


\bibitem{report} For a review of transversity see  
V.~Barone, A.~Drago and P.G.~Ratcliffe, {\it Phys. Rep.} 
\textbf{359} (2002) 1. 

\bibitem{Gasiorowicz}
 S.~Gasiorowicz, {\it Elementary Particle Physics}, 
Wiley (New York, 1967), p.~515. 
 

\bibitem{Mulders}
P.J.~Mulders and R.D.~Tangerman, {\it Nucl. Phys.} \textbf{B461} (1996)~197;
  \emph{erratum}, {\it ibid.} \textbf{B484} (1997)~538.


\bibitem{Boer1}
D.~Boer and P.J.~Mulders, {\it Phys. Rev.} \textbf{D57} (1998)~5780.


\bibitem{Boer3}
D.~Boer, P.J.~Mulders and O.V.~Teryaev, {\it Phys. Rev.} \textbf{D57}
  (1997)~3057.


\bibitem{Sivers} D. Sivers, {\it Phys. Rev.} {\bf D41} (1990) 83.

\bibitem{Mauro} M. Anselmino, M. Boglione and F. Murgia, {\it Phys. Lett.}
                {\bf B362} (1995) 164. M.~Anselmino and F.~Murgia, 
                {\it Phys. Lett.} {\bf B442} (1998) 470.

\bibitem{Boglione} M.~Boglione and P.J.~Mulders, {\it Phys. Rev.}
{\bf D60} (1999) 054007.

\bibitem{Boer2}
D.~Boer, {\it Phys. Rev.} \textbf{D60} (1999) 014012.


\bibitem{Collins}
J.C.~Collins, {\it Nucl. Phys.} \textbf{B396} (1993)~161.


\bibitem{noi}
M.~Anselmino, V.~Barone, A.~Drago and F.~Murgia, e-print hep-ph/0111044  
and paper in preparation.

\bibitem{Weinberg}
S.~Weinberg, {\it The Quantum Theory of Fields}, vol.~1, 
Cambridge University Press (Cambridge, 1995), p.~100.

\bibitem{Mauro2}
M.~Anselmino, M.~Boglione and F.~Murgia, {\it Phys. Rev.} 
{\bf D60} (1999) 054027. 


\bibitem{Bianconi}
A.~Bianconi, S.~Boffi, R.~Jakob and M.~Radici, \emph{Phys. Rev.} \textbf{D62}
  (2000) 034008.


\bibitem{Bacchetta}
A.~Bacchetta, R.~Kundu, A.~Metz and 
P.J.~Mulders, {\it Phys. Lett.} {\bf B506} (2001) 155. 


\bibitem{Jaffe}
R.L.~Jaffe, X.~Jin and J.~Tang, {\it Phys. Rev. Lett.} \textbf{80}
  (1998)~1166.


\bibitem{E704} D.L. Adams {\it et al.}, {\it Phys. Lett.} {\bf B264} (1991) 
               462.

\bibitem{Qiu}
J.-W.~Qiu and G.~Sterman, {\it Phys. Rev.} {\bf D59} (1999) 014004.

 

\bibitem{HERMES}
A.~Airapetian {\it et al.} 
(HERMES), {\it Phys. Rev. Lett.} \textbf{84}
  (2000) 4047.

\bibitem{SMC}
A.~Bravar (SMC), {\it Nucl. Phys. Proc. Suppl.} \textbf{79}
  (1999) 520.


\bibitem{desanctis}
E.~De Sanctis, W.-D.~Nowak and K.A.~Oganessyan, 
{\it Phys. Lett.} {\bf B483} (2000) 69.

\bibitem{Ma} 
B.-Q.~Ma, I.~Schmidt and J.-J.~Yang, {\it Phys. Rev.} 
{\bf D63} (2001) 037501.


\bibitem{Mauro3}
M.~Anselmino and F.~Murgia, {\it Phys. Lett.} {\bf B483} (2000) 74.





























\end{thebibliography}
\end{document}